\documentclass[draft]{agujournal2019}
\usepackage{url} %this package should fix any errors with URLs in refs.
\usepackage[inline]{trackchanges} %for better track changes. finalnew option will compile document with changes incorporated.
\usepackage{soul}
\usepackage{natbib}
\draftfalse

\journalname{Geophysical Research Letters}

\begin{document}

%%%%%%%%%%%%%%%%%%%%%%%%%%%%%%%%%%%%%%%%%%%%%%%
%  TITLE
%
% (A title should be specific, informative, and brief. Use
% abbreviations only if they are defined in the abstract. Titles that
% start with general keywords then specific terms are optimized in
% searches)
%
%%%%%%%%%%%%%%%%%%%%%%%%%%%%%%%%%%%%%%%%%%%%%%%

% Example: \title{This is a test title}

\title{Cavitons Associated with Ion-Acoustic-Like Waves in Foreshock Transients}

%%%%%%%%%%%%%%%%%%%%%%%%%%%%%%%%%%%%%%%%%%%%%%%
%
%  AUTHORS AND AFFILIATIONS
%
%%%%%%%%%%%%%%%%%%%%%%%%%%%%%%%%%%%%%%%%%%%%%%%

% Authors are individuals who have significantly contributed to the
% research and preparation of the article. Group authors are allowed, if
% each author in the group is separately identified in an appendix.)

% List authors by first name or initial followed by last name and
% separated by commas. Use \affil{} to number affiliations, and
% \thanks{} for author notes.
% Additional author notes should be indicated with \thanks{} (for
% example, for current addresses).

% Example: \authors{A. B. Author\affil{1}\thanks{Current address, Antartica}, B. C. Author\affil{2,3}, and D. E.
% Author\affil{3,4}\thanks{Also funded by Monsanto.}}

% \authors{=list all authors here=}
\authors{Runyi Liu\affil{1}, Terry Liu\affil{2}, Xin An\affil{1}, Vassilis Angelopoulos\affil{1}, and Xiaofei Shi\affil{1}}

% \affiliation{1}{First Affiliation}
% \affiliation{2}{Second Affiliation}
% \affiliation{3}{Third Affiliation}
% \affiliation{4}{Fourth Affiliation}

\affiliation{1}{Department of Earth, Planetary, and Space Sciences, University of California, Los Angeles, Los Angeles, CA, USA}
\affiliation{2}{Shandong Key Laboratory of Space Environment and Exploration Technology, Institute of Space Sciences, School of Space Science and Technology, Shandong University, Shandong, China.}
%(repeat as many times as is necessary)

% Corresponding author mailing address and e-mail address:

% (include name and email addresses of the corresponding author.  More
% than one corresponding author is allowed in this LaTeX file and for
% publication; but only one corresponding author is allowed in our
% editorial system.)

% Example: \correspondingauthor{First and Last Name}{email@address.edu}

\correspondingauthor{Runyi Liu}{runyiliu11@ucla.edu}

%%%%%%%%%%%%%%%%%%%%%%%%%%%%%%%%%%%%%%%%%%%%%%%
% KEY POINTS
%%%%%%%%%%%%%%%%%%%%%%%%%%%%%%%%%%%%%%%%%%%%%%%
%  List up to three key points (at least one is required)
%  Key Points summarize the main points and conclusions of the article
%  Each must be 140 characters or fewer with no special characters or punctuation and must be complete sentences

% Example:
% \begin{keypoints}
% \item	List up to three key points (at least one is required)
% \item	Key Points summarize the main points and conclusions of the article
% \item	Each must be 140 characters or fewer with no special characters or punctuation and must be complete sentences
% \end{keypoints}

\begin{keypoints}
\item Bursty electrostatic waves are frequently accompanied by localized electron density depletions in foreshock transients.

\item Electron density depletion scales with electrostatic potential fluctuations normalized by electron temperature.

\item Electrostatic potential provides a more robust description of wave–density coupling than electric field amplitude.

\end{keypoints}

%%%%%%%%%%%%%%%%%%%%%%%%%%%%%%%%%%%%%%%%%%%%%%%
%
%  ABSTRACT and PLAIN LANGUAGE SUMMARY
%
% A good Abstract will begin with a short description of the problem
% being addressed, briefly describe the new data or analyses, then
% briefly states the main conclusion(s) and how they are supported and
% uncertainties.

% The Plain Language Summary should be written for a broad audience,
% including journalists and the science-interested public, that will not have 
% a background in your field.
%
% A Plain Language Summary is required in GRL, JGR: Planets, JGR: Biogeosciences,
% JGR: Oceans, G-Cubed, Reviews of Geophysics, and JAMES.
% see http://sharingscience.agu.org/creating-plain-language-summary/)
%
%%%%%%%%%%%%%%%%%%%%%%%%%%%%%%%%%%%%%%%%%%%%%%%

%% \begin{abstract} starts the second page

\begin{abstract}
Foreshock transients upstream of the Earth's bow shock, such as foreshock bubbles and hot flow anomalies, are often characterized by reduced-density cores and strong plasma fluctuations. These conditions provide environments where electrostatic wave activity and localized density structures can coexist. Using high-time-resolution measurements from the Magnetospheric Multiscale (MMS) mission, we investigate the relationship between bursty electrostatic wave activity and localized electron density depletions within foreshock transients. A representative case study reveals a clear scaling between wave activity and density depletion, and a statistical analysis across multiple events shows that this scaling persists when the wave activity, with characteristics consistent with ion-acoustic-like waves, is represented in terms of electrostatic potential fluctuations normalized by electron temperature. In contrast, representations based on electric field amplitude, even when similarly normalized, exhibit substantial event-to-event variability. These results provide observational evidence for a causal relationship between ion-acoustic-like electrostatic wave activity and cavitons in foreshock plasmas.

\end{abstract}

\section*{Plain Language Summary}

Upstream of Earth’s bow shock, the solar wind can generate short-lived disturbances known as foreshock transients. These regions often contain low-density cores and intense plasma fluctuations. Understanding how plasma waves interact with these density structures helps clarify how energy is redistributed in near-Earth space. Using high-time-resolution measurements from the Magnetospheric Multiscale (MMS) mission, we examine the relationship between bursty electrostatic waves and localized reductions in electron density. We find that stronger density depletions are consistently associated with larger electrostatic potential fluctuations when scaled by the electron temperature, and this relationship holds across multiple events. In contrast, representations based on electric field amplitude show greater variability from event to event. These results suggest that ion-acoustic-like electrostatic waves play an important role in forming small-scale density cavities, known as cavitons, in the foreshock region.

%%%%%%%%%%%%%%%%%%%%%%%%%%%%%%%%%%%%%%%%%%%%%%%
%
%  BODY TEXT
%
%%%%%%%%%%%%%%%%%%%%%%%%%%%%%%%%%%%%%%%%%%%%%%%

%%% Suggested section heads:
% \section{Introduction}
%
% The main text should start with an introduction. Except for short
% manuscripts (such as comments and replies), the text should be divided
% into sections, each with its own heading.

% Headings should be sentence fragments and do not begin with a
% lowercase letter or number. Examples of good headings are:

% \section{Materials and Methods}
% Here is text on Materials and Methods.
%
% \subsection{A descriptive heading about methods}
% More about Methods.
%
% \section{Data} (Or section title might be a descriptive heading about data)
%
% \section{Results} (Or section title might be a descriptive heading about the
% results)
%
% \section{Conclusions}

\section{Introduction}
Cavitons are localized density depressions accompanied by strong, fast-oscillating electrostatic fields and represent a nonlinear manifestation of wave–plasma interactions \citep{WONGA.Y.1977C}. Their formation is commonly attributed to the ponderomotive force associated with spatial gradients in electrostatic wave intensity, which expels electrons and generates an ambipolar electric field that subsequently drives ions outward, leading to localized density depletion \citep{GoldmanM.V.1984Stop}. In addition to their role in nonlinear wave evolution, cavitons have been proposed as efficient particle accelerators: interactions with their localized electric fields can energize charged particles, and repeated interactions with propagating nonlinear structures may result in cumulative energy gain \citep{WONGA.Y.1977C}.

Upstream of Earth’s bow shock, the foreshock region frequently hosts transient plasma structures such as foreshock bubbles and hot flow anomalies, characterized by low-density, low-magnetic-field cores accompanied by significant plasma deflection and heating (e.g., \citealp{alma9985248573606533, ZhangHui2022DTPa}). These structures introduce strong spatial gradients and nonstationary plasma conditions on ion and sub-ion scales, providing favorable conditions for the generation and evolution of wave activity (e.g., \citealp{ShiXiaofei2023IWWa, GuoZhenyuan2023SPoL}).

Among these waves, ion-acoustic waves couple electron pressure perturbations to ion density fluctuations and play an important role in mediating energy transfer between electrons and ions. In collisionless space plasmas, they are frequently observed in association with particle beams and strong plasma gradients (e.g., \citealp{GoodrichKatherineA.2019IRIA, MalaspinaDavidM.2024FIAW}), making them a natural link between electrostatic wave activity and density structures in the foreshock environment.

Particle acceleration at collisionless shocks is generally understood as a multi-step process involving large-scale shock dynamics, transient structures, and wave–particle interactions in the foreshock region (e.g., \citealp{LiuTerryZ.2019Rega, ShiXiaofei2025Ceaa, RaptisSavvas2025Raul}). Within this framework, localized nonlinear structures such as cavitons may contribute to particle energization in concert with larger-scale waves and transients.

In this study, we investigate the relationship between electrostatic wave activity and localized density depletions within foreshock transients using high-time-resolution measurements from the Magnetospheric Multiscale (MMS) mission. We combine a representative case study with a statistical analysis across multiple events to examine how electrostatic wave intensity relates to density depletions and to assess the robustness of this relationship under varying plasma conditions.

\section{Dataset}
The observations used in this study are provided by the Magnetospheric Multiscale (MMS) mission \citep{BurchJ.L.2015MMOa}. Electric field measurements are obtained from the Electric Double Probe (EDP) instrument at a sampling rate up to 8192~Hz \citep{ErgunR.E.2016TADP, LindqvistP.-A.2016TSDP}. Magnetic field measurements are obtained from the Fluxgate Magnetometer (FGM) \citep{RussellC.T.2014TMMM}, and ion and electron moments are derived from the Fast Plasma Investigation (FPI) \citep{C.Pollock2016FPIf}.

Ion and electron moments derived from FPI are used to characterize the background plasma conditions, including density, temperature, and bulk flow. The temporal resolution of the FPI moments is 150 ms for ions and 30 ms for electrons, which is often insufficient to resolve localized density depletions associated with small-scale nonlinear structures such as cavitons.

To overcome this limitation, high-time-resolution electron density is inferred from the spacecraft potential using a commonly adopted exponential relationship of the form
\begin{equation}
n_e = n_0 \exp(-\phi / a),
\end{equation}
where $\phi$ is the spacecraft potential. To minimize the influence of electrostatic wave activity on the density inference, the low-pass filtered spacecraft potential with a cutoff frequency of 30 Hz is used. The parameters $a$ and $n_0$ are determined by performing a linear fit in logarithmic space between the low-pass filtered spacecraft potential and FPI electron density over a 14-s interval centered on the transient core, which is identified by the minimum electron density.

Foreshock transient intervals analyzed in this study are selected based on a previously published MMS event list \citep{LiuTerryZ.2022Ssof}. The event list contains 131 foreshock bubbles or hot flow anomalies, from 2017-11 to 2019-01. 

Within these intervals, we focus on bursty electrostatic wave activity observed in the electric field measurements. To isolate wave-like electric field fluctuations from slowly varying background fields associated with large-scale transient structures and spacecraft charging effects, a high-pass filter also above 30 Hz is applied to the electric field data. This cutoff frequency is chosen to be above the local lower-hybrid frequency while remaining well below the ion plasma frequency, thereby removing low-frequency electric field variations and retaining ion-acoustic-like wave signatures.

Previous MMS studies have identified similar large-amplitude electrostatic wave bursts in the bow shock regions as ion acoustic waves, using inter-antenna interferometry on electric field measurements to determine their propagation directions and phase speeds \citep{VaskoI.Y.2022IWia}.

A rigorous mode identification is not performed for every event included in the statistical analysis. Instead, the high-pass-filtered electrostatic wave activity is used as a proxy for ion-acoustic-like dynamics. For a subset of events, the wave properties are verified to be consistent with ion acoustic waves.

\section{Case Study: Wave Amplitude and Density Depletion}
\begin{figure}
    \centering
    \includegraphics[width=1\linewidth]{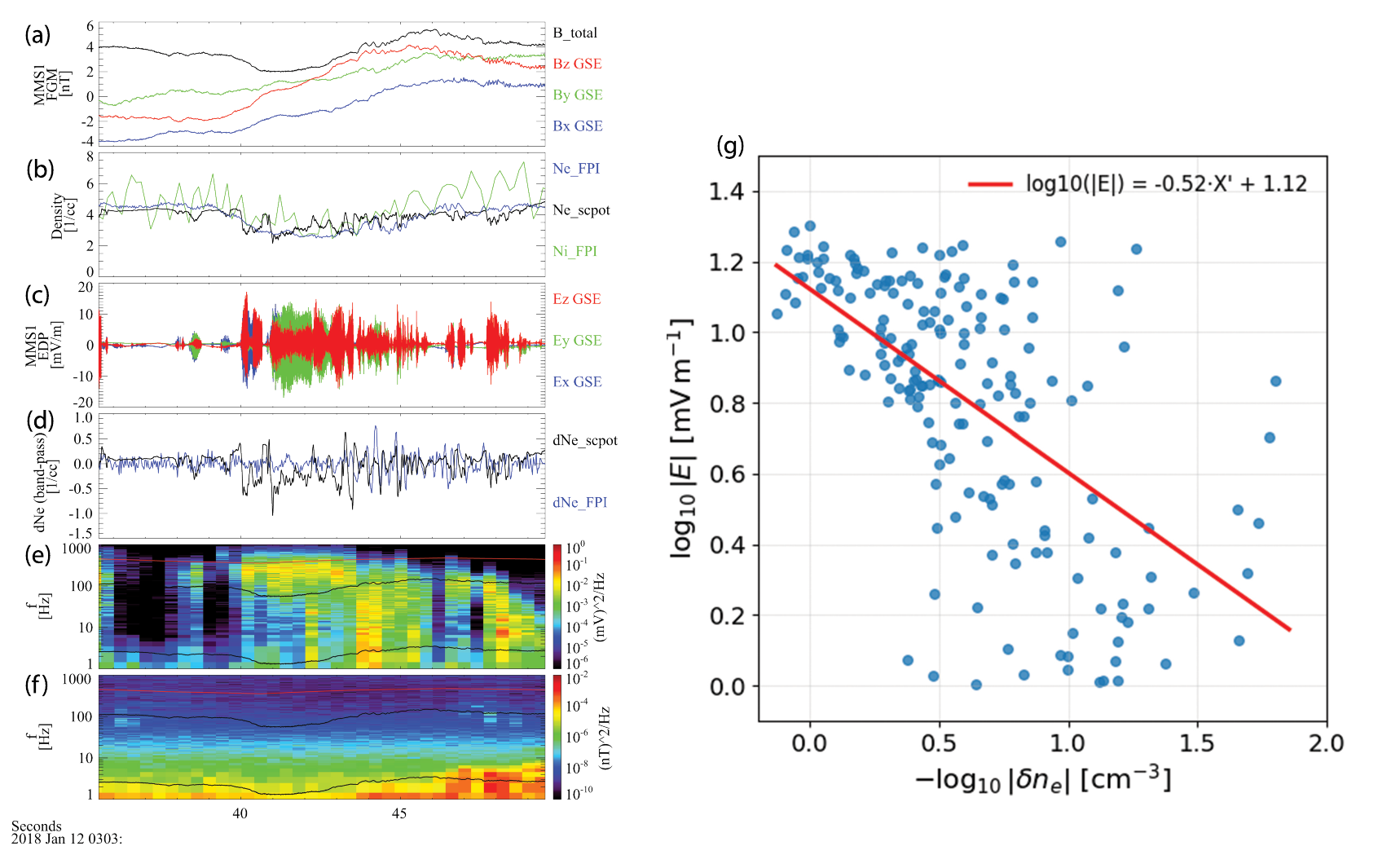}
    \caption{Overview of a representative foreshock transient observed by MMS1. (a) Magnetic field magnitude and components in the Geocentric Solar Ecliptic (GSE) coordinate system. (b) Ion and electron number densities from FPI together with the electron density inferred from the spacecraft potential. (c) Electric field components in GSE coordinates. (d) Band-pass filtered electron density (1–30 Hz) derived from the spacecraft potential, highlighting localized density depletions. (e–f) Dynamic power spectra of the electric and magnetic fields, respectively. The horizontal reference lines indicate, from low to high frequency, the lower-hybrid frequency, electron cyclotron frequency, and ion plasma frequency. (g) Scatter plot showing the relationship between the electric-field envelope amplitude and the density-depletion amplitude on logarithmic scales, with a linear fit overlaid.}
    \label{fig_1}
\end{figure}

Around 03:03:42 UT on 12 January 2018, the MMS spacecraft observed a foreshock transient. Figure \ref{fig_1} presents an event overview based on MMS1 observations to illustrate the relationship between electrostatic wave activity and localized density variations.

Figure \ref{fig_1}a shows the background magnetic field during the interval, with a pronounced reduction in field magnitude near the transient core around 03:03:42 UT. Figure \ref{fig_1}b displays plasma density measurements. Ion density moments from FPI exhibit large uncertainties and are not suitable for resolving small-scale structures. The electron density moments do show signatures of localized density depletions, but these appear weaker and less sharply defined than those inferred from the spacecraft potential, likely due to the limited temporal resolution of the FPI measurements. By contrast, the spacecraft-potential-derived electron density agrees with the FPI density on large scales while revealing pronounced localized density variations near the transient core. Several structures exhibit central density minima accompanied by enhanced density at their boundaries (e.g., around 03:03:40 UT), consistent with caviton signatures \citep[e.g.,][]{WONGA.Y.1977C}.

Figure \ref{fig_1}c shows intense bursty electric field fluctuations exceeding the background level. The band-pass filtered (1–30 Hz) electron density in Figure \ref{fig_1}d highlights localized density depletions on time scales comparable to the electric field bursts, demonstrating a clear temporal correspondence.

Figures \ref{fig_1}e and \ref{fig_1}f present the dynamic power spectra of the electric and magnetic fields. Enhanced electric field power is observed near the ion plasma frequency, with no corresponding magnetic enhancement, indicating predominantly electrostatic wave activity.

To quantify the relationship between wave activity and density depletion, we construct the scatter plot in Figure \ref{fig_1}g using envelope measures of the electric field and a depletion-only measure of the electron density.

For the electric field, we first apply the 30~Hz high-pass filter to remove slowly varying background variations. Because our goal is to characterize the wave-packet envelope rather than the individual oscillations, we compute the magnitude $|{\bf E}|$ and select its local maxima within a 50~ms sliding window. To exclude time periods without bursts, we further discard points with $|{\bf E}| < 0.05\,|{\bf E}|_{\max}$ (where $|{\bf E}|_{\max}$ is the maximum value within the analyzed interval).

For the electron density, we focus on depletion signatures only, using the band-pass filtered electron density derived from the spacecraft potential as a proxy for density variations on caviton scales. Starting from the density perturbation $\delta n_e(t)$, we remove the local positive excursions by subtracting the maximum value within a 0.5~s window. Here, $\delta n_e(t)$ denotes the band-pass filtered electron density, rather than a background-subtracted density fluctuation,
\begin{equation}
\delta n_{e,\mathrm{dep}}(t) = \delta n_e(t) - \max_{t'\in[t-0.25\,\mathrm{s},\,t+0.25\,\mathrm{s}]} \delta n_e(t').
\end{equation}
This procedure suppresses compressional enhancements and retains negative-going density depletions. The depletion amplitude is then taken as $|\delta n_{e,\mathrm{dep}}|$.

Finally, the electric-field envelope amplitude and the corresponding density-depletion amplitudes are paired and displayed on logarithmic axes, i.e., $\log_{10}(|{\bf E}|)$ versus $-\log_{10}(|\delta n_{e,\mathrm{dep}}|)$, to emphasize the dynamic range and facilitate comparison across the event.

A linear fit is applied in logarithmic space to quantify the relationship between the electric-field envelope amplitude and the density-depletion amplitude. The fitted slope is approximately $-0.5$, indicating a scaling of the form
\begin{equation}
\delta n_e \propto -|{\bf E}|^{2}.
\end{equation}
This quadratic scaling indicates that deeper density depletions are associated with stronger electrostatic wave activity and is broadly consistent with theoretical expectations \citep[e.g.,][]{GoldmanM.V.1984Stop}.

This quadratic dependence further supports a nonlinear wave–plasma interaction, in which the density response scales with wave energy density rather than linearly with wave amplitude. While this result is obtained from a single representative event, it motivates the hypothesis that a similar scaling may persist more broadly. In the following statistical analysis, we examine the robustness and prevalence of this relationship across a broader set of foreshock transient events.

\section{Statistical Study}
To assess whether the scaling identified in the case study is a general feature of foreshock transients, we apply the same analysis to all selected events in our dataset using MMS1 measurements. For each event, electric-field envelope amplitudes and density-depletion amplitudes are constructed following the procedures described above. The density perturbation is obtained using the same band-pass filter range as in the case study, with the lower cutoff set to 2~Hz based on tests between 1 and 5~Hz to suppress large-scale transient variability while retaining caviton-scale structures. In addition to the relative threshold $|\mathbf{E}| > 0.05\,|\mathbf{E}|_{\max}$, we require an absolute threshold of $|\mathbf{E}| > 0.2$~mV~m$^{-1}$ to exclude intervals dominated by the background motional electric field. The resulting pairs are combined to form the statistical ensemble.

Because plasma conditions vary between events, we introduce scaled measures to facilitate cross-event comparison. The electric field amplitude is expressed in the dimensionless form $e|\mathbf{E}|\lambda_D/T_e$, where $\lambda_D$ is the electron Debye length and $T_e$ is the background electron temperature. Density perturbations are normalized by the background electron density for each event. Background density and temperature are defined from time-resolved FPI measurements, and events are retained only if the interquartile range normalized by the median (IQR/median) of both quantities is less than 40\%, ensuring relatively stable plasma conditions.

\begin{figure}
    \centering
    \includegraphics[width=0.8\linewidth]{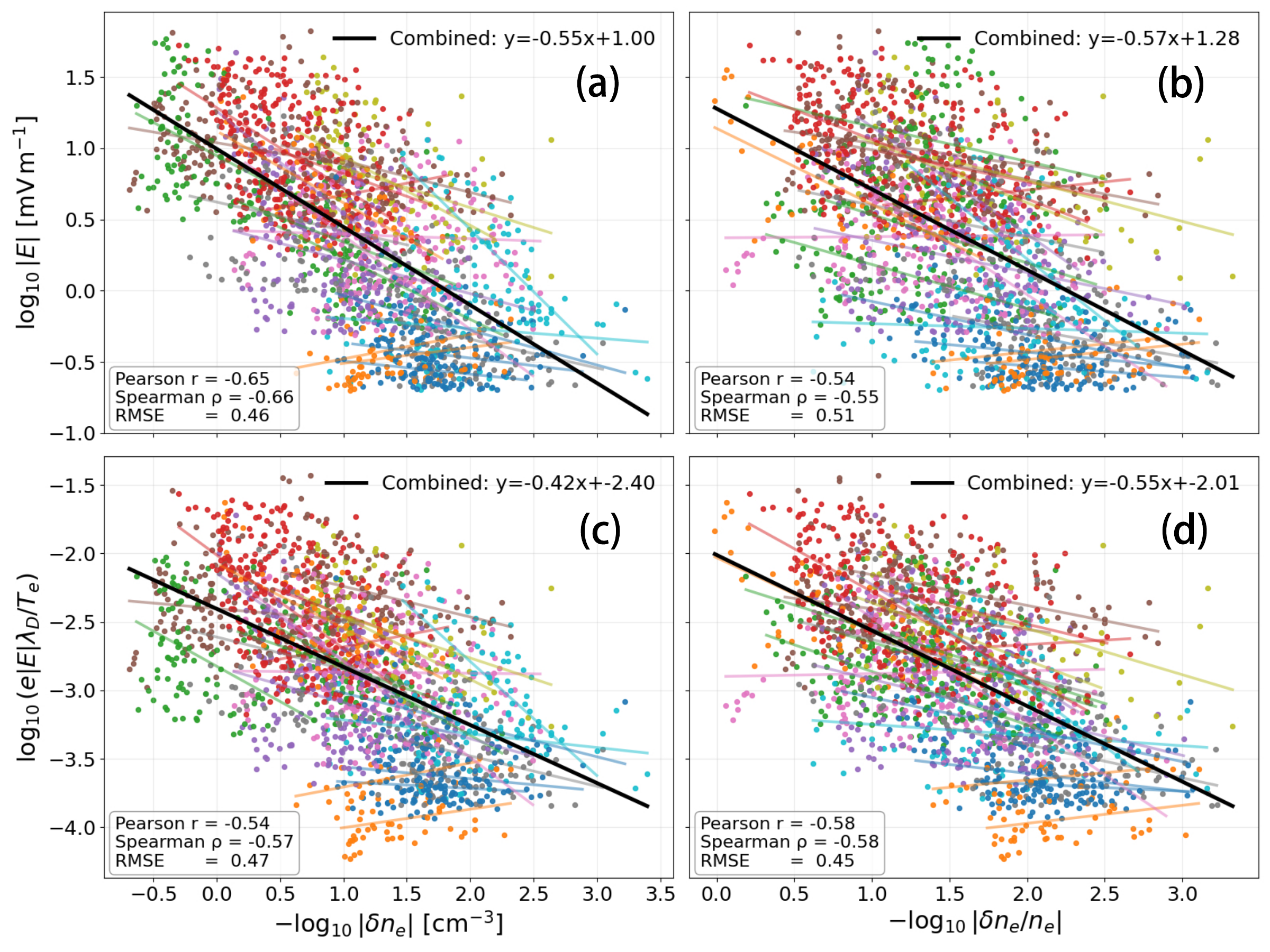}
    \caption{Statistical relationship between electric-field amplitude and density depletion for all selected foreshock transient events. Scatter plots combine data from all events using different parameterizations: (a) $\log_{10}(|\mathbf{E}|)$ versus $-\log_{10}(|\delta n_e|)$, (b) $\log_{10}(|\mathbf{E}|)$ versus $-\log_{10}(|\delta n_e/n_e|)$, (c) $\log_{10}(e|\mathbf{E}|\lambda_D/T_e)$ versus $-\log_{10}(|\delta n_e|)$, and (d) $\log_{10}(e|\mathbf{E}|\lambda_D/T_e)$ versus $-\log_{10}(|\delta n_e/n_e|)$. Here $|\mathbf{E}|$ is the envelope amplitude of the high-pass-filtered electric field, $\lambda_D$ is the electron Debye length, and $T_e$ is the background electron temperature. The quantity $|\delta n_e|$ denotes the depletion-only density amplitude derived from band-pass-filtered spacecraft-potential-based electron density. Colored symbols denote individual events, with thin colored lines showing event-wise linear fits. The thick black line indicates the linear fit to the combined dataset in logarithmic space; Pearson and Spearman correlation coefficients and the RMSE are shown in each panel.}
    \label{fig_2}
\end{figure}

Despite this physically motivated scaling and event selection, the combined scatter remains sensitive to event-to-event plasma variations (Figure \ref{fig_2}). In particular, the slopes obtained from individual events exhibit substantial variability, and the linear fit applied to the combined dataset deviates from the $\sim -0.5$ scaling identified in the case study. Moreover, both the Pearson and Spearman correlation coefficients remain moderate (of order 0.5), and the root-mean-square error of the combined fit is comparatively large. These results indicate that the scaled electric field amplitude does not provide a sufficiently robust basis for comparing wave–density relationships across different plasma environments. We therefore turn to an alternative characterization based on electrostatic potential fluctuations, which more directly capture the wave-induced electrostatic perturbation and are less sensitive to variations in characteristic spatial and temporal scales between events.

To characterize the wave activity using electrostatic potential fluctuations, we convert the electric field measurements to the corresponding potential under the assumption that the observed electrostatic wave bursts can be approximated as plane waves. Under this assumption, the electrostatic potential fluctuation $\delta \Phi$ is related to the electric field amplitude $|{\bf E}|$ through the wave number $k$.

In principle, the wave number can be directly estimated using inter-antenna interferometry, which provides wave number estimation of the waves. However, this approach is not always applicable in a statistical context, as it requires the electrostatic wave bursts to satisfy the plane-wave approximation, sufficiently large time delays between antenna signals, and high cross-correlation coefficients between opposing probe pairs. These conditions are not consistently met across all events and wave packets in the dataset.

Instead, we estimate the characteristic wave number using the observed wave frequency together with the background solar wind convection speed, assuming that the waves are convected past the spacecraft. This approach provides a wave number estimate for all events, but its accuracy depends on the relative importance of the ion acoustic phase speed compared to the background solar wind speed. Because the ion acoustic speed is not negligible at 1~AU and can reach a substantial fraction of the solar wind speed, this assumption may introduce systematic uncertainties in the estimated wave number that vary between events and wave packets.

To mitigate this effect, we examine the distribution of the estimated wave number within each event. If the IQR/median of the wave number is less than 40\%, a single representative wave number is adopted for the entire event and used to convert the electric field amplitude to the electrostatic potential fluctuation for all wave bursts within that event. Events that do not satisfy this criterion are excluded from the statistical analysis involving electrostatic potential fluctuations.

\begin{figure}
    \centering
    \includegraphics[width=0.8\linewidth]{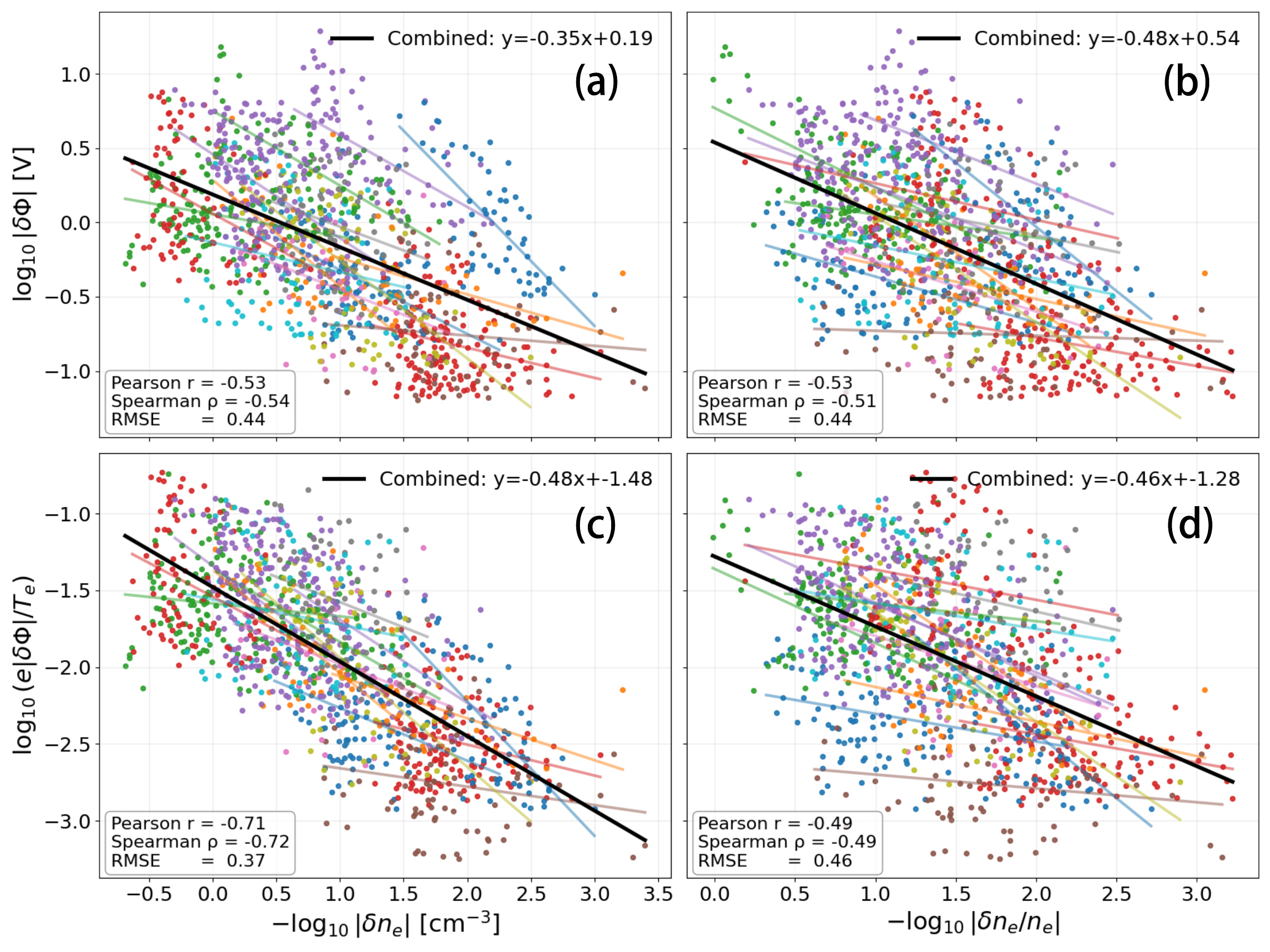}
    \caption{Same format as Figure 2, except that $|\mathbf{E}|$ is replaced by $\delta \Phi$.}
    \label{fig_3}
\end{figure}

Figure \ref{fig_3} shows the statistical scatter constructed using electrostatic potential fluctuations in place of the electric field amplitude, following the same procedures for event selection, burst identification, and density depletion extraction described above. The plotting format, binning, and fitting approach are identical to those used for the electric-field-based analysis to facilitate direct comparison.

In this representation, the electrostatic potential fluctuation $\delta \Phi$ is normalized by the background electron temperature $T_e$, while the density depletion amplitude $|\delta n_e|$ is retained without additional normalization (Figure \ref{fig_3}c). Compared to the electric-field-based scatter, this parameterization leads to reduced event-to-event variability in the event-level linear-fit slopes and yields a systematically higher correlation coefficient and a lower root-mean-square error for the combined fit across all events.

Importantly, the linear fit applied to the combined data points in logarithmic space remains close to a slope of $-0.5$, corresponding to a scaling of the form
\begin{equation}
\delta n_e \propto -\left|\frac{\delta \Phi}{T_e}\right|^{2}.
\end{equation}

Together, these results indicate that expressing the wave activity in terms of electrostatic potential fluctuations provides a more robust statistical representation of the relationship between wave activity and density depletion across different plasma environments. This parameterization both reduces event-to-event variability and preserves the scaling identified in the case study, supporting the interpretation that the observed relationship reflects a common underlying physical process.

\section{Discussion and Summary}
Foreshock transients such as foreshock bubbles and hot flow anomalies are characterized by reduced-density cores and strong plasma fluctuations. These environments provide favorable conditions for electrostatic wave activity and localized density depletions to coexist and interact, supporting the development of caviton-like structures.

Several observational limitations should be noted. Although the analysis focuses on ion-acoustic-like wave activity, the high-pass filtering applied to the electric field measurements does not uniquely isolate ion acoustic waves. Other electrostatic modes, including electrostatic components of whistler-mode waves, and electron cyclotron harmonic waves, may contribute to the observed bursty electric field signatures. In addition, the identification of localized density depletions relies on band-pass filtering of the spacecraft-potential-derived electron density, which represents a practical compromise for suppressing large-scale background variations but may not fully separate wave-induced density fluctuations from caviton-related density structures. Finally, the conversion from electric field fluctuations to electrostatic potential fluctuations relies on assumptions of electrostatic plane-wave structure and an approximate event-level wave number, introducing uncertainty into the inferred potential amplitude.

Despite these limitations, the statistical results indicate a robust relationship between electrostatic wave activity and density depletion when the wave activity is characterized using electrostatic potential fluctuations. Compared to electric-field-based scaling, which depends on background density and electron temperature through the Debye length, the potential-based representation reduces sensitivity to variations in plasma parameters across events. The combined dataset preserves a quadratic scaling between density depletion and normalized potential fluctuation, consistent with a density response governed by wave energy density rather than linearly by wave amplitude. This scaling supports the role of nonlinear processes in shaping electrostatic wave activity within foreshock transients, where bursty wave packets are often observed to evolve into more localized nonlinear structures such as electrostatic solitary waves. In this context, caviton-related density depletions may provide favorable conditions for wave localization and self-focusing. Future work will further investigate the excitation mechanisms of ion-acoustic-like waves in these environments, including the possible role of low-frequency fluctuations in generating ion beams that subsequently drive ion-acoustic-like wave activity (e.g., \citealp{WangShan2020ACSo, AnXin2024CETf, ShiXiaofei2026ETfM}).

In summary, using high-time-resolution MMS observations, we examined the relationship between bursty electrostatic wave activity and localized density depletions within foreshock transients. A representative case study and statistical analysis demonstrate that the scaling between wave activity and density depletion is most consistently recovered when the wave activity is expressed in terms of electrostatic potential fluctuations normalized by electron temperature. These results provide observational constraints on the coupling between electrostatic waves and density structures in collisionless foreshock plasmas and motivate future investigations into the role of caviton-associated structures in foreshock particle acceleration.

\acknowledgments
R. L. acknowledges support by the NASA FINESST Grant 80NSSC23K1633. X. A. acknowledges the support by NASA grants NO. 80NSSC22K1634. 

\section*{Conflict of Interest}
The authors declare no conflicts of interest relevant to this study.

\section*{Open Research}

This study was conducted using publicly available MMS data from the Fast Plasma Investigation (FPI), Electric Double Probe (EDP), and Fluxgate Magnetometer (FGM) instruments. The data are provided at the MMS Science Data Center, Laboratory for Atmospheric and Space Physics (LASP), University of Colorado Boulder, and are accessible via https://lasp.colorado.edu/mms/sdc/public/datasets. The specific time intervals analyzed in this study are described in the main text and figures.

%%%%%%%%%%%%%%%%%%%%%%%%%%%%%%%%%%%%%%%%%%%%%%%
% REFERENCES and BIBLIOGRAPHY
%
\bibliography{agusample}
% \input{agusample.bib}
% don't specify bibliographystyle
%
%%%%%%%%%%%%%%%%%%%%%%%%%%%%%%%%%%%%%%%%%%%%%%%

%\bibliography{ enter your bibtex bibliography filename here }

%Reference citation instructions and examples:
%
% Please use ONLY \cite and \citeA for reference citations.
% \cite for parenthetical references
% ...as shown in recent studies (Simpson et al., 2019)
% \citeA for in-text citations
% ...Simpson et al. (2019) have shown...
%
%
%...as shown by \citeA{jskilby}.
%...as shown by \citeA{lewin76}, \citeA{carson86}, \citeA{bartoldy02}, and \citeA{rinaldi03}.
%...has been shown \cite{jskilbye}.
%...has been shown \cite{lewin76,carson86,bartoldy02,rinaldi03}.
%... \cite <i.e.>[]{lewin76,carson86,bartoldy02,rinaldi03}.
%...has been shown by \cite <e.g.,>[and others]{lewin76}.
%
% apacite uses < > for prenotes and [ ] for postnotes
% DO NOT use other cite commands (e.g., \citet, \citep, \citeyear, \nocite, \citealp, etc.).
%

\end{document}